\magnification=\magstep1
\hsize=5.5 true in

\newcount\equationno      \equationno=0
\newtoks\chapterno \xdef\chapterno{}
\newdimen\tabledimen  \tabledimen=\hsize
%
%
%
\def\eqn{\eqno\eqname}
\def\eqname#1{\global \advance \equationno by 1 \relax
\xdef#1{{\noexpand{\rm}(\chapterno\number\equationno)}}#1}

\centerline{INITIAL DATA AND THE FINAL FATE OF INHOMOGENEOUS DUST COLLAPSE}
\vfill
\centerline{\bf I. H. Dwivedi and P. S. Joshi  }
\centerline{\bf Tata Institute of Fundamental Research}
\centerline{\bf Homi Bhabha Road, Bombay 400 005}
\centerline{\bf India}
\vfill
\noindent{\bf Proofs to be sent to:}\hfill\break
\noindent{\bf Dr. Pankaj  S. Joshi}\hfill\break
\noindent{\bf Theoretical Astrophysics Group}\hfill\break
\noindent{\bf T.I.F.R. Homi Bhabha Road}\hfill\break
\noindent{\bf Colaba, Bombay 400 005}\hfill\break
\noindent{\bf India}\hfill\break
\vfill

\vfil\eject

\magnification=\magstep1  
\hoffset=0 true cm        
\hsize=6.0 true in        
\vsize=8.5 true in        
\baselineskip=24 true pt plus 0.1 pt minus 0.1 pt 
\overfullrule=0pt         

\centerline{\bf Abstract}

We examine here the relevance of the initial state of a collapsing dust 
cloud towards determining it's final fate in the course of a continuing 
gravitational collapse. It is shown that given any arbitrary matter 
distribution $M(r)$ for the cloud at the initial epoch, there is always a 
freedom to choose rest of the initial data, namely the initial
velocities of the collapsing spherical shells, 
so that the collapse could result either in a black hole or a 
naked singularity depending on this choice. Thus, given the initial density 
profile, to achieve the desired end state of the gravitational 
collapse one has to give a suitable initial velocity to the cloud.
We also characterize here a wide 
new family of black hole solutions resulting from inhomogeneous dust collapse.
These configurations obey the usual energy conditions demanding the 
positivity of energy density.


\vfil\eject

\noindent{\bf 1. Introduction}
\bigskip

One of the important questions regarding the naked singularities arising 
from gravitational collapse is the issue of their genericity. Even if naked 
singularities occurred for reasonable forms of matter satisfying energy 
conditions etc, if this occurrence could be shown in a suitable sense to be 
non-generic, arising from an initial data set of zero measure only, then such 
a result would open up a new path for a proper formulation and possible proof 
for the cosmic censorship hypothesis which prohibits the occurrence of naked 
singularities. To make any progress towards such a goal, it is necessary to 
understand the role of initial data in determining the final fate of  
gravitationally collapsing massive objects in terms of either a black hole or 
a naked singularity. The end product of such a collapse- namely the spacetime 
singularity- is either completely covered by an event horizon, thus producing 
a black hole in the spacetime, or in the case otherwise it has possible causal
connection with the outside universe, which would require further 
investigations.

A crucial aspect of the gravitational collapse of a compact object is
the choice of initial data. After the star has exhausted its
nuclear fuel, it is well known that if the mass of the
remnant cloud was large enough, gravitational collapse would ensue.
At the time when the gravitational collapse starts,
one of the sets of initial quantities are density, pressures, etc., which
describe the state of matter distribution completely. Another set of
initial data has to be in the form of the distribution of initial
velocities of the cloud towards the center. These two sets of
initial data for the collapse would actually arise as
the end product of the process of the star exhausting its nuclear fuel.

Our purpose here is to analyze, from such a perspective, the role of 
initial data (defined in terms of physical variables such as the initial
density and velocity distributions) 
towards determining the final fate of a gravitationally 
collapsing cloud. We consider here the case of pressure-free dust collapse 
described by the Tolman-Bondi-Lemaitre models [1,2], 
which is a large class of collapse 
scenarios including the homogeneous case [3]. These models have been studied 
widely for the occurrence or otherwise of naked singularity [4-8] (for further
details and related issues, see e.g. [9]). However, the space of initial data 
which would cause a naked singularity or otherwise, 
and it's genericity, yet remain to be properly understood. 
The initial data in this case is characterized by two free 
functions on spacetime (obtained through the integration of Einstein 
equations), namely the mass function $M(r)$ describing the matter distribution,
and the velocity distribution function $V_{i}(r)$ of the spherical shells of
cloud at the onset of collapse. We study here the conditions 
causing the occurrence of a black hole or a naked singularity and it is shown 
quite generally that given any distribution of mass $M(r)$ (which is part of 
the initial data to be specified), there always exists rest of the initial 
data in the form of a suitable choice of the initial velocities $V_{i}(r)$
of the cloud towards the center, 
such that the final outcome of the collapse could be either one in the form of
a black hole or a naked singularity. 

Since the specification of initial state of matter and 
densities at the onset of collapse constitutes very important initial data 
physically, we investigate here in what way the choice of any 
particular regular distribution of matter would affect the evolution and
the final fate of the system in terms of either of these outcomes. 
For example, in the consideration by Oppenheimer and Snyder [3], the density 
distribution chosen for the (pressure free) dust cloud at the initial and also
at all later times is homogeneous (i.e. $\rho=\rho(t)$), with initial
spatial curvature being constant (i.e. initial velocity distribution
$V_i(r) \propto r$). As is well known, the collapse of such a 
configuration results in a black hole where the resulting singularity is 
fully covered by an event horizon. A generic density profile, however, need 
not be homogeneous and would be typically higher at the origin and decreasing 
away from the center. Similarly, even if the density profile is homogeneous,
the initial velocity of the spherical shells need not necessarily 
be proportional 
to $r$. It is thus essential in realistic considerations that 
the effects of both the inhomogeneities present in the initial 
matter distribution, and the variations in the velocity profiles, be 
taken into account while investigating the evolution of collapse as we shall
discuss below.  
\bigskip

\noindent {\bf 2. Dust Collapse}
\bigskip

Inhomogeneous dust collapse in the comoving coordinates (i.e. $u^i=\delta^i_t$)
is given by the Tolman-Bondi metric 
$$ds^2= -dt^2+{R'^2\over1+E}dr^2+R^2(d\theta^2+sin^2\theta d\phi^2)\eqn\qq$$
$$T^{ij}=\epsilon \delta^i_t \delta^j_t,\quad \epsilon=\epsilon(t,r)={M'
\over R^2R'}\eqn\qq$$
$$\dot R=-\sqrt{{M\over R}+E}\Rightarrow t-t_0(r)=-{R^{3/2}G(-ER/M)\over 
\sqrt{F}}\eqn\qq$$
Here $T^{ij}$ and $\epsilon$ denote the energy-momentum tensor and 
total energy density respectively.
The dot and the prime denote 
partial derivatives with respect
to the coordinates $t$ and $r$ respectively,
and we have taken 
$\dot R(t,r)<0$ as we are considering collapse scenarios only.
$G(y)$ is a strictly real positive and bound function which has the
range $1\ge y\ge - \infty$, and is given by
$$\eqalign{ G(y)&=\left ( {sin^{-1}\sqrt{y}\over y^{3/2}}-{\sqrt{1-y}\over y}
\right )\quad\hbox{for}\quad 1\ge y>0 \cr
G(y)&={2\over3}\quad \hbox{for}\quad y=0\cr
G(y)&=\left ({-\sinh^{-1}\sqrt{-y}\over (-y)^{3/2}}-{\sqrt{1-y}\over y}
\right )\quad\hbox{for}\quad 0> y\ge -\infty \cr}\eqn\qq$$
The functions $M$, $t_0(r)$ and $E$ are arbitrary functions of $r$. 
In the context of the cosmic censorship conjecture in a physically realistic
collapse scenario,  at the onset of collapse the star should have regular
initial data (density, pressure, etc.), and the spacetime be singularity free
at this initial epoch of time.
The curvature singularity would develop later during the final stages of
the  gravitational collapse. 
We therefore require that at the onset of collapse, at time $t=t_i$,
The spacetime is non-singular
and initial data in terms of density etc. do not diverge.  
Using the remaining coordinate freedom left in the choice of
coordinate $r$ we
rescale the coordinate $r$ at the
initial epoch of time $t=t_i$ as 
$$R(t_i,r)=r\eqn\qq$$
This then implies from equation (3) that $t_0(r)$, and 
the important quantity $R'$ are given by
$$ t_0(r)=t_i+{r^{3/2}G(-Er/M)\over \sqrt{M}}\eqn\qq$$
$$R'
=\left((\eta-\beta)
Y+(\Theta-(\eta-{3\over 2}\beta)
Y^{{3\over 2}}G(pY))(-p+{1
\over Y})^{1\over 2}\right)\eqn\qq$$
where we have put
$$ Y={R\over r}, \quad \eta=r{M'\over M}\quad \beta=r{E'\over E}
\quad p=p(r)=-r{E\over M}$$
$$\Theta={\sqrt {M}\over \sqrt {r}}t_0'(r)
={1+\beta-\eta \over
(1-p)^{1/2}}+(\eta -{3\over 2}\beta)G(p) 
\eqn\qq$$
The time $t=t_0(r)$ corresponds to $R=0$ which is the singularity of 
spacetime where the area of the shells of 
matter at a constant value of the coordinate $r$ vanishes, and
corresponds to the time when the matter shells
meet the physical singularity. Thus the singularity occurs at the
coordinate value $r$ at the time $t=t_0(r)>t_i$ 
and hence the range of coordinates is given by
$$0\le r < \infty, -\infty < t <t_0(r)\eqn\qq$$

From equation (3)
(i.e. $\dot R^2=E+M/R$), we
note that $M(r)$ is interpreted as the mass function for the cloud, 
and is related to the density of the matter at the onset of collapse by
$${M'\over r^2}=\epsilon (t_i,r)=\rho(r)\eqn\qq$$
The function $E(r)$ is interpreted as the energy of the cloud, 
and is related to the initial radial velocity $V_i(r)$ of the particles at
$t=t_i$ by the relation
$$E(r)= \dot R(0,r)^2-M(r)/r=V_i^2(r)-M/r\eqn\qq$$
For a free fall collapse with $\dot R(t_i)=0$, $V_i(r)=0$.
For a spherically symmetric star of radius $r_c$, the complete set of 
initial data includes the state of matter inside the star, i.e. the 
matter density $\rho(r)$, and the radial velocities $V_i(r)$
at $t=t_i$.
Note that the Kretchmann scalar $K= R^{ijkl}R_{ijkl}$ for the 
Tolman-Bondi-Lemaitre 
spacetimes is given by
$$K=12 {M'^2\over R^4R'^2}-32{MM'\over R^5R'}+48{M^2\over R^6}\eqn\qq$$
As the initial data, including the density,
should be regular and singularity free
at $t=t_i$, this implies that throughout this initial surface (i.e. for 
all allowed values of $r$), we have
$$ M(r) \equiv r^3 g(r)\eqn\qq$$
where $g(r)$ is a differentiable function. 
We assume that the weak energy condition ($\epsilon (t,r) \ge 0$) is
satisfied, and therefore $M'\ge 0$ (note that at $t=t_i$,
$\epsilon =M'/r^2$),
and consequently R is a monotone increasing function of $r$ i.e. $R'\ge 0$.
Since $M(0)=0$ (in the case otherwise, there will already be a singularity
at $r=0$ on the initial surface $t=t_i$ as seen from equation (12)), 
and due to
energy conditions $M'\ge 0$, we have $M(r)\ge 0$ for $r\ge 0$.

As we are interested in the gravitational collapse scenario of a star,
the energy density $\epsilon$  is taken to have a compact support on an initial
spacelike hypersurface, and the Tolman-Bondi-Lemaitre spacetime is matched at
some $r=const. =r_c$ to the exterior Schwarschild metric,
$$ ds^2=-(1-2M_s/r_s)dT^2+(1-2M_s/r_s)^{-1}dr_s^2 + r_s^2d\Omega^2\eqn\qq$$
The value of the 
Schwarschild radial coordinate is $r_s=R(t,r_c) $ at the boundary of the
star. We have $M(r_c)=2M_s$ where $M_s$ is the
total Schwarschild mass enclosed within the dust ball of coordinate radius 
$r=r_c$. Without going into further
details of the matching conditions, we would like to say a few words regarding
the apparent horizon. The apparent horizon in the interior dust ball
lies at $R=M(r)$. From (3) and (6) one can see that the corresponding time
$t=t_{ah}(r)$ is given by
$$t=t_{ah}(r)=t_0(r) - MG(-E)\eqn\qq$$
Since $t_{0}(r)>t_{ah}$ for all $r>0$,
and $t_0(0)=t_{ah}(0)$ at $r=0$, 
it follows that, 
only the singularity at $r=0$
could be naked, and rest of the singular points for the values $r>0$ 
are censored.

Any radial light ray terminating at this singularity at $r=0$ in the
past could go to the future infinity if it reaches the surface of the
cloud $r=r_c$ earlier than the apparent horizon at $r=r_c$. In
such a case the singularity would be globally naked.
The shell-focusing central singularity, which we discuss here, 
appears at a finite time $t=t_0(0), r=0$; however,
for $g(0)=0$ it occurs at an infinite value of coordinate time, i.e. 
$ t_0(0)=\infty$, unless
$E(r)=r^2 f_0(r)$ in such a way that $f_0(0)>0$. 
One can therefore also choose suitable radial
velocity $V_i(r)$ ( for example, $E(r)\propto r^{a},\quad a<2$)
at the initial $t=t_i$, such that for a given initial
density distribution the singularity does not form at $r=0$ (i.e.
it occurs at an infinite value of the coordinate time $t$). In such a case,
since any singularity forming at $r>0$ is covered, 
this gives rise to a black hole as the end state of collapse.

As mentioned earlier, the main aim here is to show that for a
given density distribution one can choose the initial radial velocities
of the spherical shells such that gravitational collapse could result
either in a black hole or a central naked singularity. 
We analyze here mainly the nature of the central 
shell-focusing singularities at $R=0$. However, as seen from
(2), there is a density singularity in space-time either when $R=0$, or
when $R'=0$ which corresponds to a shell-crossing singularity in the 
spacetime. A relevant question here may be regarding the occurrence of
such shell-crossing singularities, and their relevance to our conclusions. 
The point is, during the evolution of the collapse, if a shell-crossing 
singularity occurs before the formation of the physical shell-focussing
singularity $R=0$ (where the shells of matter are crushed to a vanishing
area), the metric and the coordinate system cannot be readily continued
beyond such a shell-cross 
singularity. It may be noted, however, that such shell-crosses
are generally regarded to be gravitationally weak singularities,
through which the spacetime may be continued. It is known that 
in the limit of approach to such a singularity, the spacetime curvatures
do not diverge sufficiently fast in the context of the dust models being
considered here [6], and at least a $C^1$ extension of the spacetime exists 
through such a singularity. It is also possible to discuss the extension
of spacetime in a distributional sense through such a singularity (see e.g.
[9] for a further discussion). In any case,
for any given density distribution (i.e. for given mass function $M(r)$)
we shall consider here only those energy functions $E(r)$ (i.e. the initial
velocities $V_i(r)$) for which the dust collapse 
does not give rise to any shell-crossing
singularities for $r>0$.

To make this specific, we have from equations (4) to (8)
\noindent
$$R'
={\sqrt{1-pY}\over \sqrt{Y}}\left( {Y^{{5\over 2}}G^1(pY)\over
\sqrt{1-p}G^1(p)}+\Theta(1-
{Y^{{5\over 2}}G^1(pY)\over G^1(p)})
+{\eta Y^{{3\over 2}}G(pY)\over 2}(1-{{pY
G^1(pY)\over G(pY)}\over {p
G^1(p)\over G(p)}})
\right)\eqn\qq$$
where $G^1(f)\equiv (dG(f)/df)$.
Since in a collapse scenario (i.e. $\dot R <0$), for $t\ge t_i$ 
$Y=R/r\le 1$ and because the functions $G(y)$, $G^1(y)$ are strictly 
increasing positive functions, the coefficients of
$\Theta$ and $\eta$ are positive in the above expression. 
Hence the conditions for no shell-crossings basically turn out 
to be $\Theta \ge 0$ for the case when
$\eta=(rM'/M)\ge 0$. However, for the dust cloud satisfying the weak 
energy condition and with the  mass function $M(r)\ge 0$ as pointed out 
above, we have $\eta \ge 0$. Thus, to avoid
shell-crossing singularities we must have $\Theta \ge 0$. 
As seen from (8), this amounts to
$t_0(r)$ being a monotone increasing function of $r$. 
Thus the singularity curve must be an increasing function and successive
singular points in space form at successively later times. Physically,
this means that the shells of matter with increasing values of $r$
arrive at the singularity one after the other at later and later times
without crossing each other in between. In the case of density functions
which are decreasing as one moves away from the center ( $g'(r)\le0$), 
and for the class $E\le0$, the condition
above reduces to $p'\ge0$ [10]. 

\bigskip

\noindent {\bf 3. Causal Structure and Outgoing Trajectories}
\bigskip

For a particular initial data set to develop either into a naked singularity
or a black hole, one has to analyze the behavior of causal trajectories
and categorize different situations as to when the spacetime permits outgoing
nonspacelike geodesics from the singularity, or otherwise. We therefore 
consider the radial null geodesics, given by
$$ {dt\over dr}=\pm{R'\over \sqrt{1+ E(r)}}\eqn\qq$$
where the positive sign represents outgoing solutions and negative sign 
represents ingoing ones. We consider here only outgoing geodesics, and
recall for completeness some notation and terminology from [7], which
will be used here. If there are no outgoing null trajectories escaping 
away from the   
singularity with their past end point at the singularity, then the
resulting configuration is a black hole in any case.
Therefore, for the positive sheet of solutions, we have
$${dt\over dr}={R'\over \sqrt{1+E}}
\Rightarrow {dR\over du}={1\over u'}
(R'+\dot R {dt\over dr})=(1-{\sqrt{E+\Lambda /X}
\over \sqrt{1+E}}){H}=U(X,u)\eqn\qq$$
where $u=u(r)$ is a differentiable monotone increasing function of $r$  
such that $u(r) > 0$ for $r>0$, 
$ u(0)=0$ and, from equation (7)
\noindent
$${R'\over u'}
=\left((\eta_u-\beta_u)
X+(\Theta_u-(\eta_u-{3\over 2}\beta_u)
X^{{3\over 2}}G(-PX))(P+{1
\over X})^{1\over 2}\right)\equiv H(X,u)\eqn\qq$$
$$X=(R/u),
\quad \eta_u={u\over ru'}\eta,
\quad \beta_u={u\over ru'}\beta$$
$$P=(uE/M),\quad \Lambda={M
\over u},\quad \Theta_u=\Theta{\sqrt{r}\over \sqrt{u}u'}
\eqn\qq$$

The function $\Theta_u(r)$ defined 
above plays an important role in deciding the end state of collapse. The
singularity at $R=0, u=0$ is a singularity of the differential equation (15).
For a given set of initial data (i.e. $M(r)$, and $V_i(r)$), if there exists
a function $u=u(r)$ such that $\Theta_u\sqrt{P+1/X}$ has non-zero definite
value as $r\rightarrow 0$ along all $X=$ constant directions, 
then the null geodesics could terminate in the past at the singularity, 
otherwise geodesics do not terminate there. 
The necessary and sufficient condition [7] for the central singularity 
at $R=0,r=0$ (occurring at a time $t=t_0(0)>t_i$) to have outgoing 
characteristics 
terminating in the past at the singularity then is that there should exist a 
real positive value of $X=X_0$ such that
$$X_0=\lim_{R\to 0,u\to 0}{R\over u}=\lim_{R\to 0,u\to 0}{dR\over du}=U(X_0,0)
\eqn\qq$$
If $X_0$ is positive, this implies that geodesics are outgoing with past
end point at the singularity, while for negative values of $X_0$ they are 
ingoing. Therefore, for the singularity to be naked, a positive real root of 
the following equation must exist, 
$$V(X)= U(X,0)-X=
\left(1-{\sqrt {E_0+{\Lambda_0
\over X}}\over \sqrt{1+E_0}}\right)H(X,0)-X=0\eqn\qq$$
where $E_0=E(0), \quad \Lambda_0=\Lambda(0)$. 
If there are no such positive real roots existing,
then the singularity must necessarily be covered 
in the sense that there are no outgoing null geodesics from it, 
and the collapse would result into
a black hole. On the other hand, if such a value
$X=X_0$ exists then near the singularity the characteristics
have the behavior $R=X_0u$ in the $(R,u)$ plane.
In such a case, 
since $X_0>\Lambda_0$ and $\Lambda_0$ describes the tangent to the
apparent horizon  $R=F$ at the singularity, the singularity is naked
at least locally. 

It follows that for a given initial density profile
(i.e. for a given $M(r)$),
if one could make a suitable choice of the initial velocity function, that is
$V_i(r)$, such that (22) admits real positive roots, 
the gravitational collapse would then result in at least a
locally naked singularity. The behavior of the roots equation
(22) is mainly determined by the values of
initial velocity $V_i(r)$, and its gradient $V_i'(r)$ at $ r=0$,
for a given mass function $M(r)$. Thus, a right choice of the initial
radial velocity and gradients for the spherical shells in the neighborhood
of the center $r=0$ towards the center could result in a locally naked
singularity, or otherwise. In other words, for a given density distribution
of matter, the velocity of the spherical shells in the neighborhood of the
center $r=0$ (or the energy function $E(r)$) 
at the onset of collapse determines the final state
of the collapse, ending either in a naked singularity or a black hole.

However, for a faraway external observer viewing the collapse, the singularity
would be visible if and only if there are causal trajectories with their past 
end point at the singularity, crossing the boundary  $r=r_c$ of the star 
with $dR/du >0$. An important question is, 
given a star of radius 
$r=r_c$ and total mass $M=M(r_c)$, would any singular geodesics reach the 
outside of the star to a faraway observer in the Schwarschild 
region. To be specific, we can ask for a cloud with a given value of the mass 
and size at the onset of collapse, when a locally naked singularity
will be globally visible. Our considerations 
above show that for any given initial matter 
distribution for the cloud, it is the local behavior of rest of the 
initial data, namely the corresponding energy
function $E(r)$ or the initial particle velocities, 
which decide the final fate of collapse in terms of a 
black hole, or a locally naked singularity. On the other hand, the global
visibility of the singularity will depend on the global behavior
of the corresponding functions.

To discuss this issue of global visibility, we note that 
the future behavior of outgoing escaping geodesics, which terminate in 
the past at the singularity, is determined by the global behavior of 
free functions in the form of initial radial velocity
$V_i(r)$ which appear in the geodesic equations.  In order that
the singularity be globally visible for a given density distribution $M(r)$,
the global behavior of the initial radial velocity
of the spherical shells within the cloud needs to be selected 
at the onset of collapse so that at least some geodesics reach the boundary
of the cloud before the apparent horizon with a positive value of
$dR/du$. For this purpose, we consider the paths of these null geodesics. 
From equation (15),
$${dX\over du}={1\over u}({dR\over du}-X)=
{U(X,u)-X
\over u}\eqn\qq$$
The solution of the above gives trajectories of radial null geodesics 
in the form $X=X(u)$. 
The necessary condition for a null geodesic to terminate at the
singularity at $R=0,u=0$ is that the equation 
$V(X)=0$ must have a real positive root.
Let $X=X_0$ be such a simple root of $V(X)=0$. 
By writing the above as
$${dX\over du}=
{U(X,u)-X
\over u}={(X-X_0)(h_0-1)+S\over u}\eqn\qq$$
and integrating, the null trajectories $X=X(u)$  are given by [7],
$$ X-X_0= Du^{h_0-1} +u^{h_0-1}\int Su^{-h_0}du\eqn\qq$$
where $V(X)\equiv (X-X_0)(h_0-1)+ h(X)$. The function $h(X)$ 
contains higher order terms in $X-X_0$, i.e.  $h(X_0)=(dh/dX)_{X=X_0}=0$. 
Also, 
$$h_0=1+\left[{dV(X)\over dX}\right]_{X=X_0}\eqn\qq$$
and $S=S(X,u)=U(X,u)-U(X,0)+h(X)$ is a finite and continuous function in the
region $t<t_0(r)$, i.e. for all values ($X,u$), 
and $D$ is a constant of integration that labels different geodesics.
If the singularity is the past end point of these geodesics with tangent
$X=X_0$, we must have $X\to X_0$ as $u\to 0$ in (25). Note that
as  $X\to X_0, u\to 0$, the last term in equation (25) always vanishes
near the singularity, regardless of the value of the
constant $h_0$. This is due to the reason that as $u\to 0, X\to X_0$, 
we have $S\to 0$). The first term on the
right hand side of the equation $Du^{h_0-1}$, however, vanishes only
if $h_0>1$. Therefore, the single null geodesic described by $D=0$
always terminates in the past at the singularity $R=0,u=0$, with $X=X_0$ 
as tangent. On the other hand, 
if $h_0>1$ a family of outgoing singular geodesics terminates
at the singularity in the past,
with each curve being labeled by different values
of the constant $D$. Thus the necessary and sufficient condition for
the singularity to be at least locally naked is that a real positive root
of equation (22) exist. 

In the case $h_0>1$, the integral curves given by equation (25)
do terminate at the singularity with a positive definite value of
the tangent $X=X_0$ with each null geodesic
being characterized by a value of the constant $D$, which is determined by 
the boundary conditions at $r=r_c$. For a
singular geodesic reaching the boundary of the dust cloud $u=u_c=u(r_c)$ 
with $X=(R_c/u_c)=X_c$ we have,
$$ X_c-X_0= Du_c^{h_0-1} +u_c^{h_0-1}\int_{u_c} Su^{-h_0}du\eqn\qq$$
and hence the equation of such a geodesic can be written as
$$ X-X_0= (X_c-X_0)({u\over u_c})^{h_0-1}
+u^{h_0-1}\int_{u_c}^{u} Su^{-h_0}du\eqn\qq$$
The event horizon is represented by the null geodesic for which 
$X_c=\Lambda(r_c)$. For an outgoing null geodesic terminating 
in the past at the singularity, we have $X=X_0>\Lambda_0$ as the tangent at 
the singularity, and the curve is ejected into the region
$R>F$ where $dR/du$ is positive. Therefore, all the geodesics that
reach the line $r=r_c$ (where the metric (1) is matched with the 
Schwarschild exterior) with $X_c>\Lambda (r_c)$ (note from equation (18) that
$dR/du$ for such geodesics at $r=r_c$ is positive)
would escape to infinity, while others would become
ingoing. Such geodesics from the singularity, that reach future infinity,
are then given by the equation (28) with $X_c>\Lambda_c$.

An important subcase to note is when the
root equation $V(X)=0$ has only two real positive roots $X=X_{\pm}$. In such a 
case,  $h_0-1>0$  either at $X=X_+$ or at $X=X_-$. If 
$h_0-1>0$ at $X=X_-$ then $h_0-1<0$ at $X=X_+$. In this scenario the
geodesics would be terminating at the singularity with tangent $X=X_-$, i.e.
$X\rightarrow X_-, u\rightarrow 0$. From equation (23),
the behavior of these
geodesics near $X=X_+$, assuming that $p, \eta, \beta$  are
finite as $r \to \infty$, is given by
$$(X-X_+)^{1\over 1-h_0}\propto {1\over u}\eqn\qq$$  
and thus as $X\to X_+$ the null 
geodesics attain arbitrarily large values of $r$, making the
singularity globally visible.

It follows from the above considerations that the occurrence of a 
black hole, or a locally or globally naked singularity,
would generally depend on both the local as well as global behavior of the
functions $M(r)$ and $V_i(r)$. In particular, the occurrence of a 
naked singularity (which may be locally or globally visible),
for a prechosen distribution of matter, is determined completely
by the initial velocity of the spherical shells towards the center 
at the onset of collapse. In the next section we examine the evolution 
of gravitational collapse from such a perspective.

\bigskip

\noindent{\bf 4. Black Holes and Naked Singularities}
\bigskip

Before examining the general inhomogeneous dust collapse, we first
examine the special subclass of homogeneous models, where the density
distribution at any given instant of time has a constant value in space. 
The Oppenheimer-Snyder models [3] of gravitational collapse fall
within this class, where the initial density
$ \rho=\epsilon(t_i,r)=M'/r^2=\rho_0= const.$ everywhere in space, that is 
$$M(r)=M_0r^3\eqn\qq$$
In their original paper, Oppenheimer and Snyder considered
the above constant density distribution, with the specific choice of the
initial velocity for the cloud given by,
$$ V_i(r)=-V_or,\quad V_o=const.\eqn\qq$$
It is well-known that such a set of initial data leads to a black hole.

For any given constant density profile, however, the final result of 
the collapse
could be different for different choices of initial velocity profiles as
we shall discuss here. For example, if one chooses the initial 
velocity differently so that 
$V_i(r)\propto r\sqrt{1+ar^3},\quad a=const.$, or in terms of the 
function $E(r)$ as
$$E(r)=-M_0r^2(e_0+e_1r^3+\gamma_0(r)r^{3})$$ 
where $e_0, e_1,$ are constants 
and $r^3\gamma_0(r)$ is a $C^2$ function of $r$ such $\gamma_0
(0)=0$, then such a choice of $E(r)$ leads to a 
naked singularity in gravitational collapse (see also [7]). The 
constants $e_0$ and $e_1$ are chosen here such that
$$\Gamma_0={3e_1\over e_0M_0^{{3\over 2}}}
({1\over \sqrt{1-e_0}}-{3G(e_0)\over 2})= 3e_1{G^1(e_0)\over M_0^{3/2}}> 
13+{15\over 2}\sqrt{3}\eqn\qq$$
With such a choice, $u=M_0r^3$ and the root equation (19) becomes 
$$V(X)=0\Rightarrow 2x^4+x^3-\Gamma_0(x-1)=0,
\quad X= x^2\eqn\qq$$
For the value of $\Gamma_0$ chosen as above, 
this equation has only two real positive roots, 
namely $X_{\pm}=x_{\pm}^2>1$, such
that $x_{+}>x_{-}$, which implies $X_{+}>X_{-}$.
The exact value of these roots depend on the
exact value of $\Gamma_0$.
It follows that $h_0>1$ for the root $X=X_{-}$, and hence a family of outgoing
null geodesics terminate at the singularity with the tangent $X=X_{-}$.
Also, $h_0<1$ for the other root $X=X_+$. 
The behaviour of the geodesics is given by equation (29).
To give a specific example, one can take $\Gamma_0=28$, and then the 
roots are $X_+=2.443, X_-=1.512$, and $h_0-1=0.768$ along the root $X=X_-$ and
$h_0-1= -0.433$ along the root $X=X_+$. The
boundary of the cloud lies at $r=r_c$ (note that $r_c>> 2M_s\Rightarrow
1>> M_0r_c^2$). 
In the neighborhood of the singularity
the trajectories are described by 
$$ X-X_-= (X_c-X_-)({u\over u_c})^{h_0-1}\eqn\qq$$
For the trajectories with $X_c<X_-$, $X$ decreases as $u$ increases, and they
move towards the apparent horizon which is a straight line given by $X=1$
in $(R,u)$ plane. For geodesics with $X_c>X_-$, the
value of $X$ starts increasing as $u$ increases and they move away
from the apparent horizon. 
The geodesic with $X_c=1$ crosses the boundary
of the cloud at the intersection of this boundary 
with the apparent horizon. The trajectories
with $X_c>1$ reach the boundary at an earlier time than
the apparent horizon and the singularity is globally naked. 

We note in the above case of homogeneous dust collapse 
that the quantities $\Theta$ and $p$ are given by 
$\Theta=r^2G^1(p)p'$, and $p=e_0+r^3(e_1+\gamma_0(r))$.
Therefore, in order that  shell-crossing singularities do not
occur during the collapse, it
is sufficient to require that $p'\ge 0$, or that $p$ be a monotone
increasing function of $r$. This implies 
that while fixing the energy function $E(r)$, 
the choice of $\gamma_0(r)$ be such
that $3r^2(e_1+\gamma_0(r))+r^3\gamma_0'(r)\ge 0$ for $r_c\ge r >0$.

It follows from the considerations above 
that within the framework of homogeneous dust collapse
itself, there is a wide variety of new collapse solutions which result
either into a black hole or a naked singularity, depending on the 
existence or otherwise of the real positive roots of the algebraic
equation (33). While it is well-known that the Oppenheimer-Snyder
homogeneous dust collapse with a specific choice of velocity profile
results into a black hole, we have seen here that a wide variety of 
different choices of particle velocity profiles would also produce the 
Schwarzschild black hole as the end state of gravitational collapse.

We now consider the future development of general inhomogeneous density 
profiles during the collapse. As pointed out
earlier, for the initial data to be regular at the onset of collapse
the most general mass function has to be of the form $M(r)=r^3g(r)$, 
where $g(r)$ is a differentiable function. For physical
reasonableness, we first consider the cases where the density is a decreasing
function of $r$ away from the center (hence  $g(0)>0$), and 
as such $g'(r)\le 0$.  If
$$lim_{r\to 0} \left[{r^3\over g(r)-g(0)}\right]=0\eqn\qq$$
that means $g(r)$ contains terms that are lower then $r^3$. In 
such a case, we select the energy function as that corresponding to the 
marginally bound case $E(r)=0$, and then $u$ is given by 
$$u=g^{{-1\over 3}}-g_0^{{-1\over 3}},\quad g_0=g(0)\eqn\qq$$
We then have
$$ V(X)\equiv X-{1\over \sqrt{X}}=0 \Rightarrow X=1\eqn\qq$$
Hence the singularity is naked and the geodesics terminate at the
singularity with the tangent $X=1$ in the $(R,u)$ plane. 
Since $E=0$ and $g'(r)\le0$ 
therefore there are no shell-crossing singularities
within the cloud of boundary at $r=r_c$.
On the other hand,
if equation (35) is not satisfied that means $g(r)$ contains terms
of the order $r^3$ or higher. In that case, we choose the velocity profile
as given by
$$E(r)=-e_0r^2g(r)(1+\gamma(r)r^3)\eqn\qq$$
which implies $p(r)=e_0(1+r^3\gamma(r))$, 
where $\gamma(r)$ is a differentiable function such that
$e_1=\gamma(0)> 0$ and $3+r\gamma'(r)\ge 0$. Then $u=r^3$, and
the constants $e_0$ and $e_1$ are chosen such that 
$$\Gamma_1={3e_1\over e_0M_0^{{3\over 2}}}
({1\over \sqrt{1-e_0}}-{3G(e_0)\over 2})> 13+{15\over 2}\sqrt{3}\eqn\qq$$
where $M_0=g(0)$. With such a choice the algebraic equation in
question reduces to
$$V(X)=0\Rightarrow  2x^4+x^3-\Gamma_1(x-1)=0,
\quad X={M_0} x^2\eqn\qq$$
This has two real positive roots, 
namely $X_{\pm}=M_0x_{\pm}^2$, such
that $x_{+}>x_{-}\Rightarrow X_{+}>X_{-}$
and therefore the singularity is naked. For this class under consideration,
since $g'(r)\le 0 $, we have $p'=e_0r^2(3+r\gamma'(r))\ge 0$ 
and the shell-crossings again do not occur.
Note that while the only $\gamma(0)$ and $e_0$ values determine the nakedness
of the singularity, it is the global behavior of $\gamma(r)$ for 
the range $r_c\ge r>0$ which determines that no shell-crossings occur. 

In the above discussion of evolution of inhomogeneous density
profiles, motivated by considerations of physical reasonableness, we 
assumed decreasing density profiles away from center with $g'(r)\le0$ and
$g(0)\ne 0$. However, we can consider the case of a general mass
function $M(r)= r^3g(r)$ without these assumptions. We can then choose the 
velocity of the collapsing shells as described by the energy function
$E(r)$ given by
$$E(r)=-r^2g(r)p(r),\quad G(p)= \sqrt{g}\gamma_1(u)\eqn\qq$$
where $u=F(r)=r^3g(r)$, and $\gamma_1(u)$ is a monotone increasing
function of $u$  with $\gamma_1(u)\ge 0$ and 
$$\left[{d\gamma_1\over du}\right]_{u=0}=C_0>0\eqn\qq$$
We than have
$$\Theta_u={d\gamma_1\over du}=\gamma_1,_u(u),\quad \gamma_1,_u(0)=C_0\eqn\qq$$
The equation $V(x)=0$ now becomes
$$V(X)=(1-\sqrt{{1\over X}})\left({X+{3C_0\over \sqrt{X}}\over 3}
\right)-X=0$$
$$\Rightarrow 2x^4+x^3-3C_0(x-1)=0$$
This is the same equation as (33), and would have real positive roots
if $3C_0>13+{15\over 2}\sqrt{3}$ in which case the singularity would be naked,
otherwise it would be covered. Note that for the naked singularity to occur,
the above inequality is essential and it involves only $C_0$ which
is the first derivative of $\gamma_1(u)$ at $u=0$. 
The behavior of $\gamma_1(u)$ within the cloud $r_c\ge r>0$ is chosen
as per the model one wants and its global behavior within the 
cloud is taken such that there are no shell-crossing singularities. 
The dust models would be bounded, marginally bound or unbounded ($p>0,=0,<0$) 
depending
upon the choice of $\sqrt{g}\gamma_1 >{2\over 3}, ={2\over 3}, <{2\over 3}$ 
respectively. 

For nonoccurrence of shell-crossing singularities, 
first note that energy conditions imply that $\eta =rM'/M>0$ for $r>0$ 
and since $\gamma_1$ is monotone increasing function, we have
$$\Theta=\gamma_1,_{u}M'r\sqrt{g}\ge 0\eqn\qq$$
Therefore, within the cloud for $r_c\ge r>0$
there are no shell-crossing singularities during the evolution of 
gravitational collapse.

It thus follows that given any density profile
for the cloud at the initial epoch, a suitable regular choice of the velocity
distribution would make the collapse of the cloud result into either 
a black hole or a naked singularity, depending on this choice.
This characterizes a wide family of new collapse solutions for 
inhomogeneous dust which result into a black hole, apart from the 
well-known black hole solution resulting from the homogeneous dust
collapse.

Although we have given above only particular initial velocity distributions
at the onset of the collapse for a given density profile, it must be
pointed out that many more similar velocity profiles would exist which 
would also ensure the fate of the collapse in terms of 
either a naked singularity or a black hole as desired. 
We have given here a treatment which includes all possible initial
density profiles, while various specific density profiles have been worked
out earlier by several authors (see e.g. [9]) for the occurrence of a naked 
singularity or black hole.

An interesting choice of a velocity distribution is worth pointing out
in the context of the discussion here. 
The null geodesics in the spacetime are given by the differential 
equation (18). The equation
involves two free functions, namely $M(r)$ and $E(r)$. Thus, for a
given $M(r)$ the exact behavior of the trajectories of photons, which are
the null geodesics) is described by the way $E(r)$ is chosen.  Consider
for example the mass function of the type given above $M(r)=r^3g(r)$, 
and let us choose $E(r)$ which satisfies 
$$ U(a,u)=a,\quad a>1$$
where $a$ is a constant. 
In general, $U$ defined by equation (18) is an explicit function of
$E, M$ and $R$ and an implicit function of $r$; choosing $R=au$ and $M=u$
gives the above equation. For such an $E(r)$
the null geodesic characterized by the equation $R=aM$ terminates at the
singularity. This geodesic never gets inside the apparent horizon
and furthermore along the trajectory we have $R'>0$ for $r>0$,
thus avoiding the shell-crossing curve also. 
The geodesic crosses the boundary of the cloud
avoiding shell-crossings (if there are any), and makes the 
singularity globally naked.

We will now consider some further details on the issue of global nakedness
for the naked singularity formation discussed above.
In reality, for an external observer viewing the collapse, the singularity
would be naked if and only if there are causal geodesics (with their past 
end point at the singularity) crossing the boundary of the star $r=r_c$ 
with $dR/du >0$, that is they cross the boundary of the star before
the apparent horizon. 
In the above discussion, we have given several models for an arbitrary
density distribution where the singularity is naked. Do every example
of a locally naked singularity, under all circumstances, 
lead to global nakedness? The answer is certainly in the negative.  
As such local nakedness already implies that for a set of given 
$M(r)$ and $E(r)$, which leads to a naked singular spacetime, causal 
geodesics which terminate at the singularity in the past do escape into 
the region $R>M$ near the singularity and as such have $dR/du$ 
positive in a finite neighborhood
of the singularity, however small. It follows that one can always choose
the boundary at $r=r_c>0$, however small, so that in this region and at
the boundary we have $dR/du>0$. Therefore one can always 
select the total mass $M(r_c)$, and the size $r_c$ of the cloud, such that the 
escaping null geodesics would reach  $r_c$  
with $dR/du$ positive for such examples, making
the singularity globally naked. Similarly one can, by choosing the 
boundary suitably, make it globally invisible in these cases.
Thus for a given set of density and velocity distributions,
which leads to a locally naked singularity, one can always choose the
the set of total Schwarschild mass $M(r_c)$, and the size of the
cloud $r_c$  so as to make the collapse either globally naked or otherwise.

For an arbitrary set of mass $M$ and size $r_c$, let us consider the 
general mass function and the example given by equation (41) 
where $E(r)=-r^2g(r)p(r)$,
$\quad G(p)= \sqrt{g}\gamma_1(u),\quad u=M$. For the physically 
reasonable case $g(0)\ne 0$ one can
take the bound case $p\le 0$. In this case, between the cloud
$r_c\ge r>0$ the only condition one has on the function $\gamma_1$
is that it be a monotone increasing function, 
i.e. $\gamma_1'=\gamma_1,_uM'\ge0$ with $\gamma_1(0)>0$. 
At the center at $r=0$ one should have
$\gamma_1,_u(0) =C_0\ge 13+{15\over 2}\sqrt{3}$, and then
the family of geodesics terminate at the singularity. The root equation (22)
has only two real positive roots $X_{\pm},\quad X_+>X_-$. 
The geodesics terminate
at the singularity with the tangent $X=X_->1$ in $(R,u)$ plane
and attain arbitrarily large
value of $r$ at $X=X_+$. In $(R,u)$ plane the apparent horizon
is a straight line $R=u$ and $X_{\pm}>1$ are the roots of equation (22).
Geodesics with $X_c>X_-$ will escape to the boundary with
$dR/du$ positive and the singularity would be globally naked. 
In cases we considered with  $g(0)=0$, the choice of $E(r)$ corresponds
only to the unbound class, however,
the conditions on $\gamma_1$ remain the same for no shell-crossings as
well as for the singularity to be naked.

So far we have concentrated on initial matter distributions of the star.
One may also ask whether the converse is true, that is, given a regular 
initial data in the form of $E(r)$, whether there always exists
a regular $M(r)$ which could produce either one of the black hole or 
a naked singularity. For example, the collapse initiating from rest
with $V_i(r)=0$ (i.e $\dot R=0$ initially) falls under such a situation
and can be considered similarly by making a suitable choice of the mass
function $M(r)$. The answer in general seems to be in affirmative, 
and one can develop the arguments along similar lines as above, though we will
not go into the details here. Another point to note is  regarding the
curvature strength of the naked singularities considered here. Not all the
examples considered here need be that of strong curvature type, except the
ones where the root equation admits two real positive roots, which will be
necessarily strong. These details will be given elsewhere [11].

\bigskip

\noindent{\bf 5. Concluding Remarks}

A relevant question here is whether results 
such as above could be generalized to more general equations of state.
In fact, recent work [12] has indicated that the phenomena of naked singularity
or black hole formation in collapse is closely related to the specification 
of initial data. It has been pointed out there that if a naked singular 
spacetime (or a black hole) 
exists for a particular equation of state then for all sufficiently 
close equations of state there would also be a naked singular solution 
(black hole) in 
spherically symmetric collapse. Hence similar conclusions as above 
should hold for such more general equations of state as well. 
Further details on this and related issues 
will be discussed elsewhere.

We have shown here that for gravitational collapse of dust
clouds with arbitrary initial matter distributions, the final fate of 
collapse in terms of a black hole or a naked singularity is fully 
determined by the choice of rest of the initial data. It follows that 
given any regular matter distribution, its collapse can always lead to the 
formation of a black hole provided the remaining initial data is chosen 
suitably; and that such a choice is always possible. In other words, given 
the initial density distribution, we have given here a specific procedure 
to fine tune rest of the initial data so as to produce a black hole as the
end product of gravitational collapse. Such a conclusion no longer needs the 
assumption of cosmic censorship conjecture. Usually, in black hole physics, 
the truth of cosmic censorship is always assumed in the sense that collapse 
from regular initial conditions is necessarily taken to be a black hole. 
However, no proof or a mathematically suitable formulation of such a 
hypothesis is available so far. We have provided here a more specific 
characterization of initial data for the formation of black holes without 
such an assumption. Thus, these considerations bring out a wide new class of 
collapse solutions producing a black hole 
from the inhomogeneous collapse of matter. We also 
pointed out initial data configurations resulting into naked singularities 
for any given distribution of matter. In view of such a situation, it appears 
important to identify and determine physically the possible initial data sets 
responsible for either of these phenomena. 

\vfil\eject

\centerline{REFERENCES}

\item{\bf [1]} R. C. Tolman, Proc. Natl. Acad. Sci. USA, 20, p.169, (1934).

\item{\bf [2]} H. Bondi, Mon. Not. Astron. Soc. 107, p.343 (1947).

\item{\bf [3]} J. R. Oppenheimer and H. Snyder, Phys. Rev. 56, p.455, (1939).

\item{\bf [4]} D. Eardley and L. Smarr, Phys. Rev. D19, p.2239, (1979).

\item{\bf [5]} D. Christodoulou, Commun. Math. Phys. 93, p.171, (1984).

\item{\bf [6]} R. P. A. C. Newman, Class. Quant. Grav. 3, p.527, (1986).

\item{\bf [7]} P. S. Joshi and I. H. Dwivedi, Phys. Rev. D47, p.5357 (1993).

\item{\bf [8]} P. S. Joshi and T. P. Singh, Phys. Rev. D, 
T. P. Singh and P. S. Joshi, Class. Quant. Grav. March (1996).

\item{\bf [9]} P. S. Joshi, `Global Aspects in Gravitation and
Cosmology', Clarendon Press, Oxford, (1993).

\item{\bf [10]} C. Hellaby and K. Lake, Ap. J. 290, p. 381 (1985);
R. P. A. C. Newman, Ref 6 above.

\item{\bf [11]} S. Jhingan and P. S. Joshi, TIFR preprint (1996).

\item{\bf [12]} P. S. Joshi and I. H. Dwivedi, Commun. Math. Phys. 
146, p.333 (1992); Lett. Math. Phys. 27, p.235 (1993); Commun. Math. Phys.
166, p.117 (1994);  See also Ref [9].
\end